# Tunneling spectroscopy of gate-induced superconductivity in MoS$_2$

Davide Costanzo[1] *, Haijing Zhang[1] *, Bojja Aditya Reddy[1], Helmuth Berger[2] and Alberto F. Morpurgo[1] **.

[1]DQMP and GAP, Université de Genève, 24 quai Ernest Ansermet, Geneva CH-1211, Switzerland.

[2]Institut de Physique de la Matière Complexe, Ecole Polytechnique Federale de Lausanne, Lausanne CH-1015, Switzerland.

*These two authors contributed equally to this work

**e-mail: alberto.morpurgo@unige.ch

**The ability to gate-induce superconductivity by electrostatic charge accumulation is a recent breakthrough in physics and nano-electronics. With the exception of LaAlO$_3$/SrTiO$_3$ interfaces, experiments on gate-induced superconductors have been largely confined to resistance measurements, which provide very limited information about the superconducting state. Here, we explore gate-induced superconductivity in MoS$_2$ by performing tunneling spectroscopy to determine the energy-dependent density of states (DOS) for different levels of electron density $n$. In the superconducting state, the DOS is strongly suppressed at energy smaller than the gap, $\Delta$, which is maximum ($\Delta \sim 2$ meV) for $n$ of $\sim 10^{14}$ cm$^{-2}$ and decreases monotonously for larger $n$. A perpendicular magnetic field $B$ generates states at $E < \Delta$ that fill the gap, but a 20% DOS suppression of superconducting origin unexpectedly persists much above the transport critical field. Conversely, an in-plane field up to 10 T leaves the DOS entirely unchanged. Our measurements exclude that the superconducting state in MoS$_2$ is fully gapped and reveal the presence of a DOS that**

**vanishes linearly with energy, the explanation of which requires going beyond a conventional, purely phonon-driven Bardeen-Cooper-Schrieffer mechanism.**

The observation of a transition to a zero-resistance state signals the occurrence of superconductivity but provides virtually no information about the microscopic nature of the superconducting state, whose determination is complex and requires the combination of multiple experimental probes. Among these, the use of tunneling spectroscopy to probe the energy-dependent density of states (DOS) has traditionally played a central role. For instance, shortly after the appearance of the BCS paper on superconductivity[1], Giaever's measurements on Al, Sn and In tunnel junctions[2,3] gave a clear experimental validation of the theory. Analogous experiments[4,5] on Pb and Nb tunnel junctions demonstrated that the attractive interaction leading to electron pairing is phonon-mediated, and enabled the quantitative determination of the electron-phonon coupling strength[6]. More recently, tunneling spectroscopy has been systematically performed on high-$T_c$ transition metal oxides[7] and iron pnictides[8,9] to explore their unconventional non *s*-wave nature. Here, we apply tunneling spectroscopy to investigate gate-induced superconductivity in the electron accumulation layer at the surface of $MoS_2$[10].

Starting with $LaAlO_3/SrTiO_3$ interfaces[11,12], the number of recently discovered material systems that allow superconductivity to be gate-induced has been steadily increasing[10,13,14,15,16,17,18,19,20]. Several of these systems are van der Waals bonded layered compounds in which superconductivity has been induced using ionic liquid gated transistors[10,14,16,17,18,19,20], among which $MoS_2$ has attracted particular attention (see Fig. 1a). For instance, superconductivity persists even when the MoS$_2$ thickness is reduced down to an individual monolayer[19] and is extremely robust against the application of an in-plane magnetic field[21,22]. Also, irrespective of the MoS$_2$ thickness, superconductivity always occurs in the two-dimensional regime because the depth of the accumulation layer is limited by the electrostatic screening length ~ 1 nm[21,22]. However, no tunneling experiments have been performed so far because the accumulation layer is buried between the liquid gate and the MoS$_2$ crystal and is difficult to access with a tunnel-coupled probe.

**Heterostructure devices for tunneling spectroscopy**

To perform tunneling spectroscopy of gate-induced superconductivity, we exploit nano-fabricated devices based on van der Waals heterostructures[23], in the configuration schematically illustrated in Fig. 1b. The devices employ a graphene multilayer (typically 6-8 layer thick) as tunneling electrode, with the exfoliated MoS$_2$ crystals (typically 6-10 layer thick) both hosting the gate-induced accumulation layer and acting as tunnel barrier. The ionic liquid covers the MoS$_2$ multilayer as well as the gate consisting of a large area metallic pad, so that a positive gate bias results in the accumulation of electrons onto the MoS$_2$ top surface. Under this condition, the band profile across the thickness of the MoS$_2$ multilayer is schematically shown in Fig. 1c. Different device configurations have been tested –by covering the edges of MoS$_2$ with a PMMA layer (Fig. 1d top) or leaving them exposed (Fig. 1d bottom)– to exclude that edge effects determine transport from the graphene multilayer to the accumulation layer (see supplementary information for more details).

The key advantage of the configuration shown in Fig. 1b is that it allows us to measure in-plane transport and to extract the DOS on a same device. The energy-resolved DOS is obtained from the voltage-dependent differential conductance measured by biasing one of the graphene strips relative to a contact attached to the top of the $MoS_2$ layer. At low-temperature, the current injected "vertically" from a graphene electrode into the accumulation layer at the surface of $MoS_2$ is carried by electrons tunneling through the un-doped part of the $MoS_2$ multilayer (see Fig. 1c). That $MoS_2$ multilayers can act as reliable tunnel barriers has been shown recently[24], and here it is indicated by the temperature dependence of both the non-linear *I-V* curves (Fig. 1e) and of the zero-bias differential resistance (Fig. 1f). As *T* is lowered from room temperature, these quantities initially exhibit a thermally activated dependence (due to thermally assisted transport), followed by a saturation as *T* is decreased below ~70 K, confirming that at low temperature "vertical" transport is dominated by tunneling. In all devices studied saturation was seen to reproducibly occur between 60 and 100 K; the observed reproducibility suggests that this value is determined by the alignment of the Fermi level in the accumulation layer and the bottom of the conduction band in the undoped part of $MoS_2$.

**Measuring the DOS in the superconducting state**

At the lowest temperature of our measurements (either 1.5 K or 250 mK, depending on the cryostat used), the "vertical" differential conductance exhibits a pronounced low-energy suppression, whose magnitude decreases as the temperature is increased, and disappears for $T \gtrsim T_c$, the critical temperature determined by the resistive transition (defined as the temperature for which the measured resistance is 95% of the normal state value). This behavior, originating from the suppressed DOS in the superconducting state, is illustrated in Fig. 2 with data from two devices gate-biased at different values of electron density $n$ (inferred from Hall resistance measurements). Specifically, Fig. 2a shows the temperature dependence of the in-plane resistance at $n = 1.5 \cdot 10^{14}$ cm$^{-2}$ from which $T_c$ is determined to be approximately 10 K, and Fig. 2b shows the normalized bias-dependent tunneling conductance recorded for different values of $T \leq 10\ K$ (the tunneling conductance is normalized by dividing it by the tunneling conductance measured at $T$ just above $T_c$ –see the inset of Fig. 2a and 2c). Analogous measurements performed at $n = 5.1 \cdot 10^{14}$ cm$^{-2}$, shown in Fig. 2c and d, exhibit a virtually identical behavior, albeit with $T_c \cong 3.5$ K.

We measured the tunneling conductance of 5 devices cooled down multiple times with different applied gate voltage, to investigate accumulation layers with electron density in the range between $n \sim 1 \cdot 10^{14}$ cm$^{-2}$ and $n \sim 5 \cdot 10^{14}$ cm$^{-2}$. The measurements exhibit common features, already visible in Fig. 2b and d. The low-energy suppression in the DOS –observed in all devices investigated– is not complete and its energy dependence is not as pronounced as it would be expected from a fully gapped system. "Side-lobes" are typically seen at finite bias (at ~1.8 mV in Fig. 2b and ~0.3 mV in Fig. 2d), with an amplitude much less pronounced than the coherence peaks expected for conventional *s*-wave superconductors. In a few cases a double lobe structure was seen (as in Fig. 2d). Finally, any time measurements were performed at $T \gtrsim 0.2\ T_c$, kinks at sub-gap energy, symmetric around zero bias, could be clearly identified, as the ones pointed to by the purple arrows in Fig. 2b and 2d.

These features result from a systematic behavior that we identify by plotting quantities measured on devices gate-biased at different values of electron density $n$, as a function of reduced variables ($eV/\Delta$ and $T/T_c$). Fig. 3a shows the normalized tunneling conductance of devices covering the full range of electron densities, measured at $T/T_c \cong 0.2$ (since $T_c$ is different for different devices, the actual value of $T/T_c$ may deviate from 0.2 by about 10-15%). It is clearly apparent that for $eV/\Delta \cong 1$ large sample-to-sample fluctuations are present, with the amplitude of the "lobes" that can vary from very pronounced to nearly absent. For $eV/\Delta < 0.5$, however, all curves exhibit an approximately linear dependence and virtually collapse on top of each other. Such a systematic low-bias behavior insensitive to the electron density $n$ is confirmed by the $V = 0$ V differential conductance plotted as a function of $T/T_c$ (see Fig. 3b). Data points measured at all values of electron density follow a common trend, which for small $T/T_c$ is linear, with the differential conductance extrapolating to zero for $T/T_c \to 0$.

The observed systematic trends rule out that the low-energy (i.e., $E < 0.5\,\Delta$) tunneling conductance is strongly affected by inhomogeneity in electron density, whose presence is virtually unavoidable in ionic liquid gated transistor at low temperature[18][19]. Indeed, measurements of the Hall resistance between different pairs of contacts on a same device show that the inhomogeneity in electron density is strongly device-dependent (see supplementary information). Therefore, any physical quantity whose behavior is critically determined by this inhomogeneity should also exhibit strong sample-to-sample fluctuations. Fig. 3a shows that sample-to-sample fluctuations affect the shape of the differential conductance for $eV \sim \Delta$ (how pronounced the side lobes are, or whether one or two lobes are seen, depends on the sample) but not at low-bias. This conclusion is further supported by the common trend followed by the $V = 0$ tunneling conductance of different devices shown in Fig. 3b, which confirms that charge density inhomogeneity does not have a dominant influence on the measured low-energy DOS. Finally, note that finding that the tunneling conductance extrapolates to zero for $T/T_c \to 0$ (see Fig. 3b) implies that the vast majority of MoS$_2$ is superconducting. This is an important conclusion because the non-vanishing resistance that is seen in some of the devices for $T \ll T_c$ (see, e.g., Fig. 2c) may be taken as an indication that a large fraction of the accumulation layer remains in the normal state. The large suppression in the low-temperature tunneling conductance seen in Fig. 3b directly shows that this is not the case (see supplementary information for a more detailed discussion).

Fig. 3c shows the evolution of $T_c$ and $\Delta$ with increasing $n$. The dependence of $T_c$ extracted from the resistive transitions connects smoothly with data reported earlier by Iwasa's group[10], which reached up to $n = 1.5 \cdot 10^{14}$ cm$^{-2}$, and extends the study of gate-induced superconductivity to $n > 5 \cdot 10^{14}$ cm$^{-2}$. In this yet unexplored regime $T_c$ decreases upon increasing electron density, indeed forming a well-defined dome. Finding a smooth connection with Iwasa's group data is noteworthy, because it demonstrates quantitative consistency of independent observations of a same phenomenon, something that had not yet been reported for ionic liquid gated superconductivity (note also that finding a systematic evolution of $T_c$ with $n$ implies again that it is the average density that determines the critical temperature, and not the density fluctuations due to the inhomogeneity present in ionic liquid gated devices, consistently with the considerations made above). From the differential conductance we also extract $T_c$ (which we define as the temperature at which a $V = 0$ suppression becomes visible beyond the noise) and the gap $\Delta$. Considering that the precise value of $T_c$ depends somewhat on the definition used, the agreement between the $T_c$ values extracted from the differential conductance and from the resistive transitions is satisfactory, and shows the consistency of transport and tunneling data. The dependence of $\Delta$ on $n$ follows that of $T_c$, albeit with larger fluctuations (see Fig. 3c - right axis). This is because the differential conductance at larger bias (i.e., $eV \sim \Delta$), from which the value of $\Delta$ is obtained, is more strongly affected by inhomogeneity in electron density (see Fig. 3a).

**Influence of magnetic field**

We discuss the evolution of the tunneling conductance in a magnetic field $B$, starting from the case of field applied perpendicularly to the MoS$_2$ surface. The tunneling conductance of two different devices measured at increasing values of $B$ is shown in Fig. 4a –with Fig. 4b showing the corresponding resistive transition– and Fig. 4c (the critical temperature and electron density for the two devices are respectively $T_c \sim 10$ K with $n = 1.5 \cdot 10^{14}$ cm$^{-2}$, and $T_c \sim 9$ K with $n = 1.8 \cdot 10^{14}$ cm$^{-2}$). In the $T_c \sim 10$ K device, the magnetic field mainly causes the appearance of low-energy states, affecting weakly the energy scale below which the DOS is suppressed –i.e., the gap (see Fig. 4a). We ascribe this effect to the vortices created in the superconductor, binding states at sub-gap energy [25]. In the second device with lower $T_c$, the generation of low-energy states is accompanied by a more apparent gap suppression (see Fig. 4c). Irrespective of these details, a large suppression in the low-energy DOS persists in both cases well above the critical magnetic field inferred from the resistive transition ($B_c$, the value of $B$ for which the resistance is 95% of the value measured at high field). For instance, for the T$_c$ ~10K device, the critical magnetic field extracted from the resistive transition is approximately 9 T, but a large suppression in the low-bias tunneling conductance is still clearly visible at the largest field used in these measurements, $B = 10$ T. For the $T_c \sim 9$ K device, $B_c$ is approximately 7 T, and a suppression in the low-energy tunneling conductance is still clearly seen at 14 T. The large-$B$ suppression in the DOS disappears when $T$ is raised above $T_c$ (see Fig. 4d), demonstrating that the phenomenon is of superconducting origin.

This unexpected behavior was observed in all devices investigated. The results are summarized in Fig. 4e by the plot of the relative suppression in the DOS as a function of $B/B_c$, which shows that a ~20% effect persists even when $B \gg B_c$. In the presence of a perpendicular magnetic field, therefore, a resistive state occurs in which superconducting correlations are nevertheless still present. This directly implies that the magnetic field at which the resistive transition occurs does not correspond to the upper critical field $H_{c2}$ beyond which superconductivity is completely suppressed, as assumed in past work[10][21][22]. It follows that the value of superconducting coherence length $\xi$ extracted by interpreting the data using Ginzburg-Landau theory[26] –i.e., the relation $H_{c2} = \frac{\Phi_0}{2\pi\xi^2}$ – is not correct. If this relation is valid at all, our data would indicate that $\xi$ is much shorter than anticipated. However, since our observations are not expected within a conventional Ginzburg-Landau scheme, it will be essential to first understand the microscopic nature of the observed phenomenon –with coexisting finite resistivity and superconducting correlations– before making any quantitative estimate. Examples of possible scenarios that should be considered to understand our observations are the quantum metal state recently discussed in thin NbSe$_2$ crystals[27], or motion of flux quanta causing the superconducting phase stiffness to vanish[26].

When the magnetic field is applied parallel to the MoS$_2$ surface, the DOS remains unchanged. This is shown in Fig. 4f with data taken at T= 1.5 K, on one of the devices discussed earlier (see Fig. 2a and Fig. 4a), upon increasing the in-plane field from 0 to 10 T (the maximum in-plane field applied in our experiments). The insensitivity of the DOS is quite striking: it can be understood that the in-plane field does not affect the orbital degrees of freedom –since the accumulation layer is only 1.5 nm thick[21][22] – but changes could still occur because *B* couples to the spin, and the Zeeman energy at 10 T is comparable to $\Delta$. The phenomenon may be attributed to the strong spin-orbit interaction (approximately 10 meV in the conduction band of MoS$_2$[28]), which forces the spin to point perpendicular to the MoS$_2$ and cause a so-called Ising superconducting state [21][22][29][30] (i.e., to influence the DOS, the Zeeman energy has to be comparable to the spin-orbit energy, which is significantly larger than $\Delta$). So far, experimental evidence for Ising superconductivity was inferred from the small effect that an in-plane field has on the critical temperature, and from the correspondingly extrapolated very large in-plane critical field at low temperature[21][22] (~ 60-80 T depending on the model chosen to extrapolate). Tunneling spectroscopy, however, provides more information because it shows directly that the in-plane magnetic field does not change the DOS anywhere on the Fermi surface.

**Conclusions**

The results of our tunneling spectroscopy measurements provide new information on the nature of the gate-induced superconducting state. In particular, finding that the low-bias (*eV/Δ* < 0.5) tunneling conductance measurements coincide when plotted as a function of *eV/Δ* irrespective of the value of $n$ (Fig. 3a), and that at $V = 0$ V all data falls on a same curve when plotted versus $T/T_c$ (Fig. 3b), show that the superconducting state is the same throughout the density range explored. Additionally, the dependence of the tunneling conductance on $V$ and $T$ consistently indicates that the low-energy DOS is linearly proportional to $E$, the distance in energy from $E_F$, and vanishes at $E = 0$. Such a linear dependence is incompatible with the presence of a full gap, which would cause a much more pronounced sub-gap suppression of the tunneling conductance over a broader bias range [26] (in the supplementary information we show that attempts to fit the data starting from a fully gapped DOS are unsuccessful). The absence of a full gap excludes several potential candidate states, such as s-wave superconductivity and an equal-sign two-band superconducting state (the same realized in $MgB_2$[31][32][33] and possibly in $NbSe_2$[31][34][35], whose relevance for gated $MoS_2$ may be suggested by the double gap structure observed in several devices; see Fig. 2d). Indeed, both these superconducting states possess a complete gap, with vanishing sub-gap DOS, even in the presence of considerable density of (non-magnetic) impurities [26].

The observed low-energy DOS suggests that understanding the gate-induced superconducting state in $MoS_2$ will require going beyond a conventional, purely phonon driven BCS mechanism. More work will be needed, and it is useful to look at theoretical studies analyzing gate-induced superconductivity in mono or bilayers of $MoS_2$ (and in other semiconducting transition metal dichalcogenides) that have recently appeared in the literature [29][36][37][38][39][40], to see if any of the proposed superconducting states could be taken as promising candidates. In this context, we find particularly interesting the proposed possibility of a so-called s± state[36] (a two-band superconducting state with opposite sign of the order parameter in each band, thoroughly investigated in the context of iron pnictide superconductors [41,42,43,44]), or of a nodal phase that may occur in the presence of interactions favoring spin-triplet pairing if the pocket at the M-Γ line on the conduction band of $MoS_2$ is populated [29]. In the appropriate regimes, these states could account for at least some of our experimental observations, such as the linear energy dependence of the DOS for $E \ll \Delta$, the presence of kinks inside the gap, and the double-gap structure that we observed in some of the devices. Whether experimental conditions are compatible with a description in terms of either one of these states, however, is not yet clear and remains to be established. Irrespective of these considerations, what is really worth emphasizing is that our results demonstrate a strategy to perform tunneling spectroscopy on a gate-induced superconductor, and show a wealth of interesting new information not accessible in conventional transport measurements. The relevance of this result goes beyond the specific case of $MoS_2$, as the experimental strategies demonstrated here can be applied to explore the gate-induced superconducting states recently discovered to occur in many other layered two-dimensional materials.

**Acknowledgements**

We gratefully acknowledge A. Ferreira for continued technical support of the experiments. We are also very grateful to Prof. K.T. Law for extended, extremely useful discussions. Financial support from the Swiss National Science Foundation, the NCCR QSIT and the EU Graphene Flagship project is also very gratefully acknowledged.

**Author contributions**

D.C., H.Z., and B.A.R. fabricated the devices, and performed the electrical measurements. D.C. and H.Z. analyzed the data. H.B. provided high quality $MoS_2$ crystals. A.F.M. conceived the experiment, directed the research and wrote the manuscript. All authors read the manuscript and gave comments.

**Competing financial interests**

The authors declare no competing financial interests.

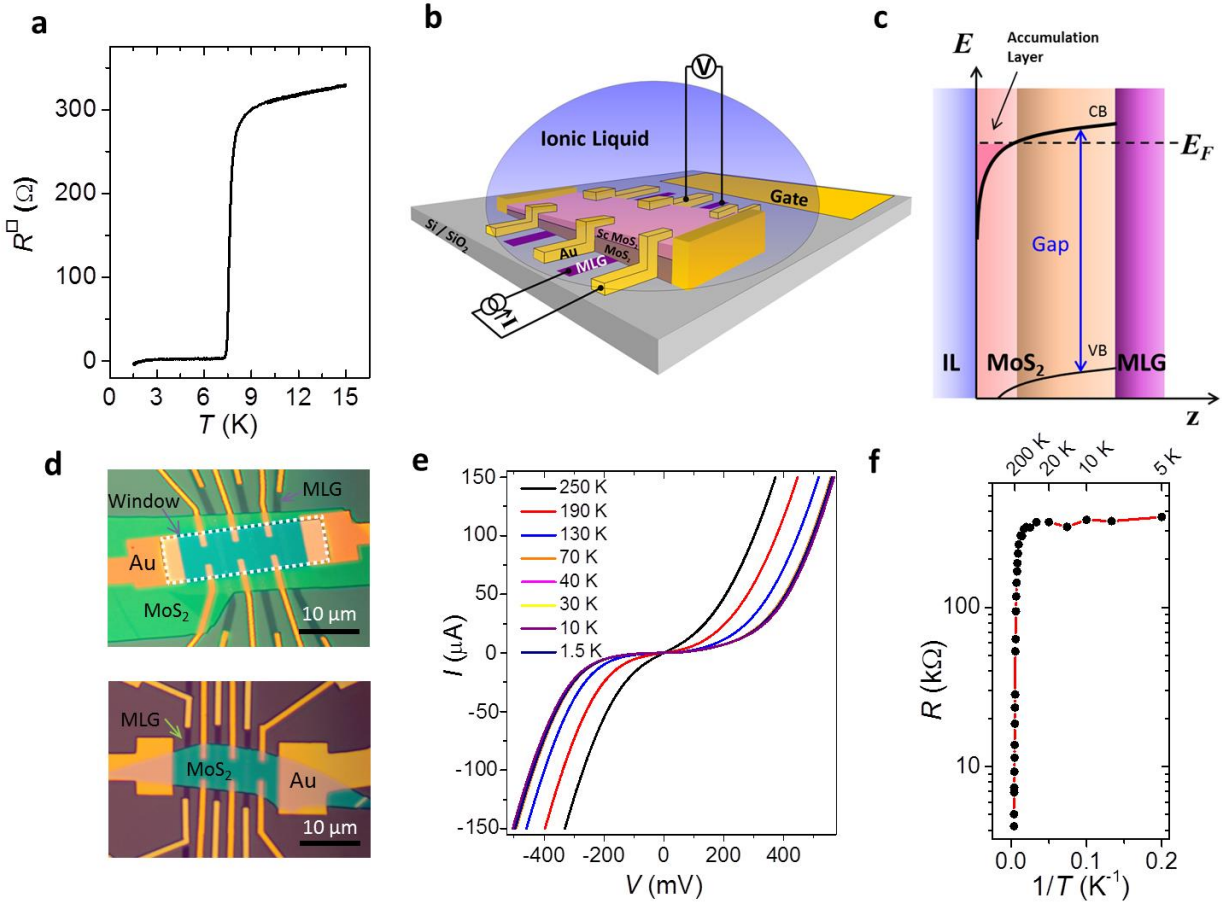

**Figure 1. Nano-fabricated MoS$_2$ devices for tunneling spectroscopy measurements**. **(a)** Example of a resistive transition demonstrating the occurrence of gate-induced superconductivity at the surface of MoS$_2$. **(b)** Schematics of the devices used to perform tunneling spectroscopy. Multilayer graphene (MLG) strips on a Si/SiO$_2$ substrate act as tunneling electrodes. An exfoliated MoS$_2$ crystal (typically 6-10 layer thick) is positioned across the strips and contacted with gold electrodes. The ionic liquid covers the structure, as well as a large Au pad acting as gate. When current is sent from a MLG strip to one of the gold contacts, the undoped part of MoS$_2$ (shaded brown in the figure) acts as a barrier and electrons tunnel from MLG to the gate-induced accumulation layer at the surface of MoS$_2$ (shaded pink in the figure). **(c)** Band-diagram across the device thickness. The ionic liquid (IL) gate induced superconducting accumulation layer extends approximately 1-1.5 nm in depth; deeper in MoS$_2$ the Fermi level is inside the gap, not far from the conduction band edge. **(d)** Optical microscope images of two devices (the bar is 10 µm). In the device on top, a layer of PMMA is used to cover the edges of the flake and define the active part of the device (region delimited by the white dashed line). The bottom device is not covered with PMMA. **(e)** The "Vertical" I-V curves, measured between a MLG electrode and one of the Au contacts, become $T$-independent below 70 K, indicating that transport is mediated by direct tunneling; also, the $V$=0 vertical resistance, thermally activated at high $T$, saturates for $T$< 70 K **(f)**.

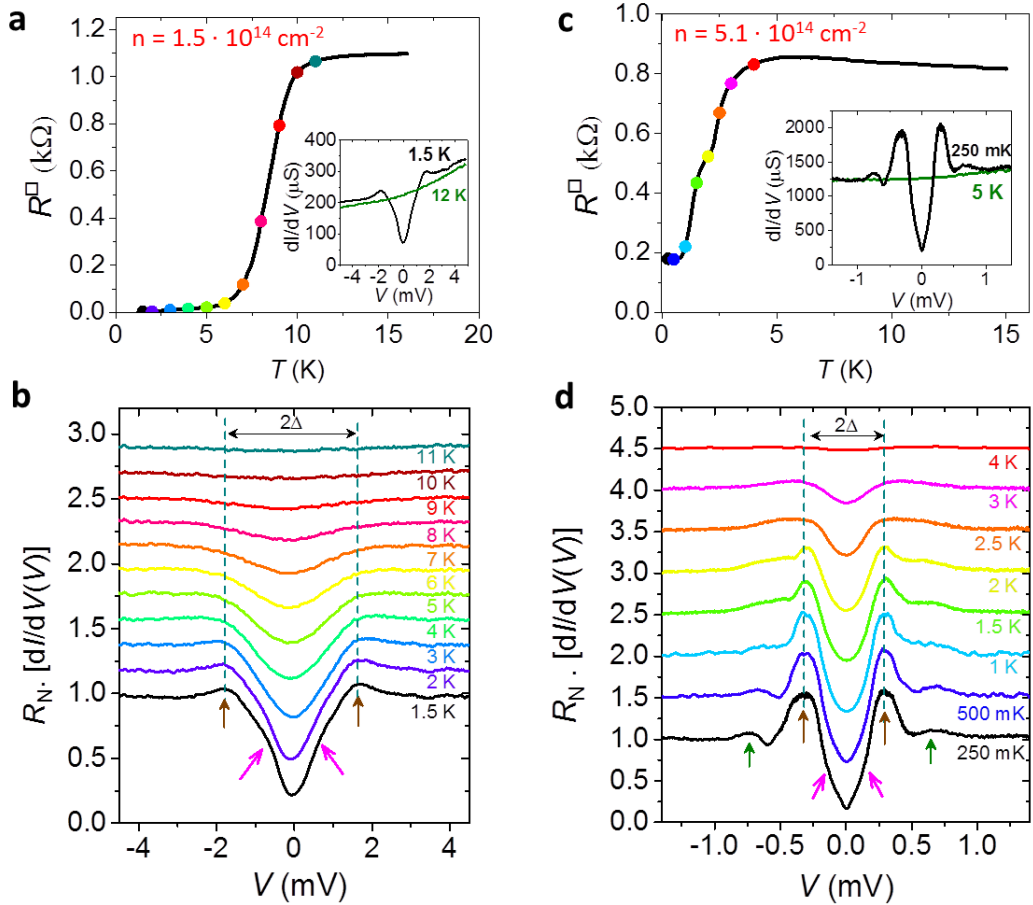

**Figure 2. Temperature evolution of the bias-dependent tunneling conductance.** Resistive transition **(a)** and temperature evolution of the normalized tunneling conductance **(b)** measured on a MoS$_2$ accumulation layer with $n = 1.5 \cdot 10^{14}$ cm$^{-2}$. The colored dots in **(a)** indicate the temperatures at which the $dI/dV$-$V$ curves of the corresponding color in **(b)** are measured. Panels **(c)** and **(d)** show the same measurements on another device, gate-biased at $n = 5.1 \cdot 10^{14}$ cm$^{-2}$ (in **(b)** and **(d)** curves are offset for clarity). In all devices, a measurable suppression of the tunneling conductance starts to be observed in correspondence of the onset of the resistive transition. Upon cooling, the suppression increases, and typically reaches 70-90 % of the high-bias value, depending on the device and on the temperature reached in the measurements. The magnitude of the suppression does not correlate with the quality of the resistive transition (for instance, in **(c)** no $R=0$ Ω state is attained, but the suppression in $dI/dV$ is more pronounced than in **(b)**). Side lobes are seen in nearly all the measured devices (pointed to by the brown arrows in **(b)** and **(d)**) and are used to determine the value of the gap $\Delta$ (see the vertical dashed lines in **(b)** and **(d)**). In all devices, a sub-gap kink in the $dI/dV$-$V$ curve is clearly visible whenever $T < 0.2\, T_c$ (pointed to by the purple arrows in **(b)** and **(d)**). In some devices a second peak is visible which may correspond to a second superconducting gap (pointed to by the green arrows in **(d)**). The insets in **(a)** and **(c)** show the raw $dI/dV$ data measured below and above $T_c$ (the normalized $dI/dV$-$V$ curves are obtained by taking their ratio).

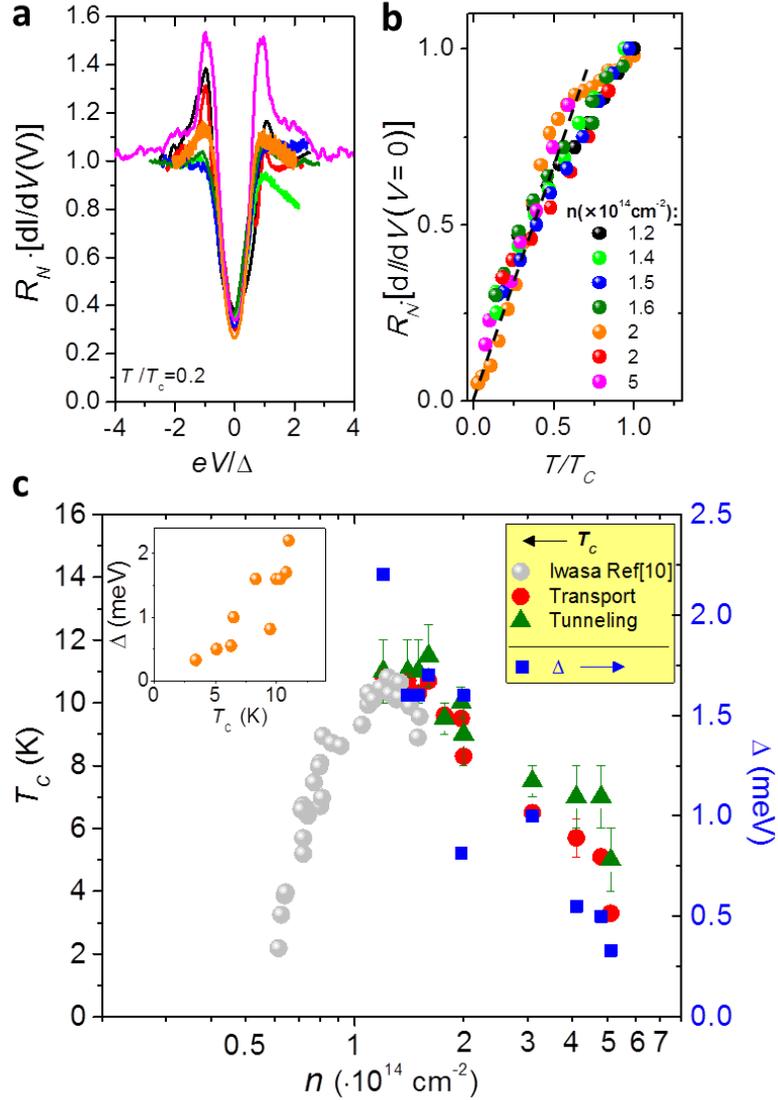

**Figure 3. Investigating the nature of the superconducting state in $MoS_2$.** **(a)** Normalized tunneling conductance as a function of $eV/\Delta$, measured at the same value of $T/T_c$ = 0.2 for different values of electron density $n$ in the range 1.2-5.1·$10^{14}$ $cm^{-2}$: all curves coincide at low bias irrespective of $n$. Also, throughout the entire range of investigated density, the $V$=0 V tunneling conductance plotted versus $T/T_c$ exhibits a common trend **(b)**, with all data falling on a same curve that extrapolates linearly to zero for small $T$. The dependence of $T_c$ and $\Delta$ on $n$ is shown in (**c**). The grey and red dots represent $T_c$ obtained from resistive transitions in Ref. [10] and in our work, respectively. The green triangles represent $T_c$ obtained from the tunneling conductance, defined as the temperature for which the DOS suppression exceeds the experimental noise. The blue squares represent the superconducting gap $\Delta$ (see right axis) determined from the position of the side lobes in the tunneling conductance measured at $kT \ll \Delta$ (vertical dashed lines in Fig. 2b and d). The inset shows the correlation between $\Delta$ and $T_c$ obtained from the experimental data.

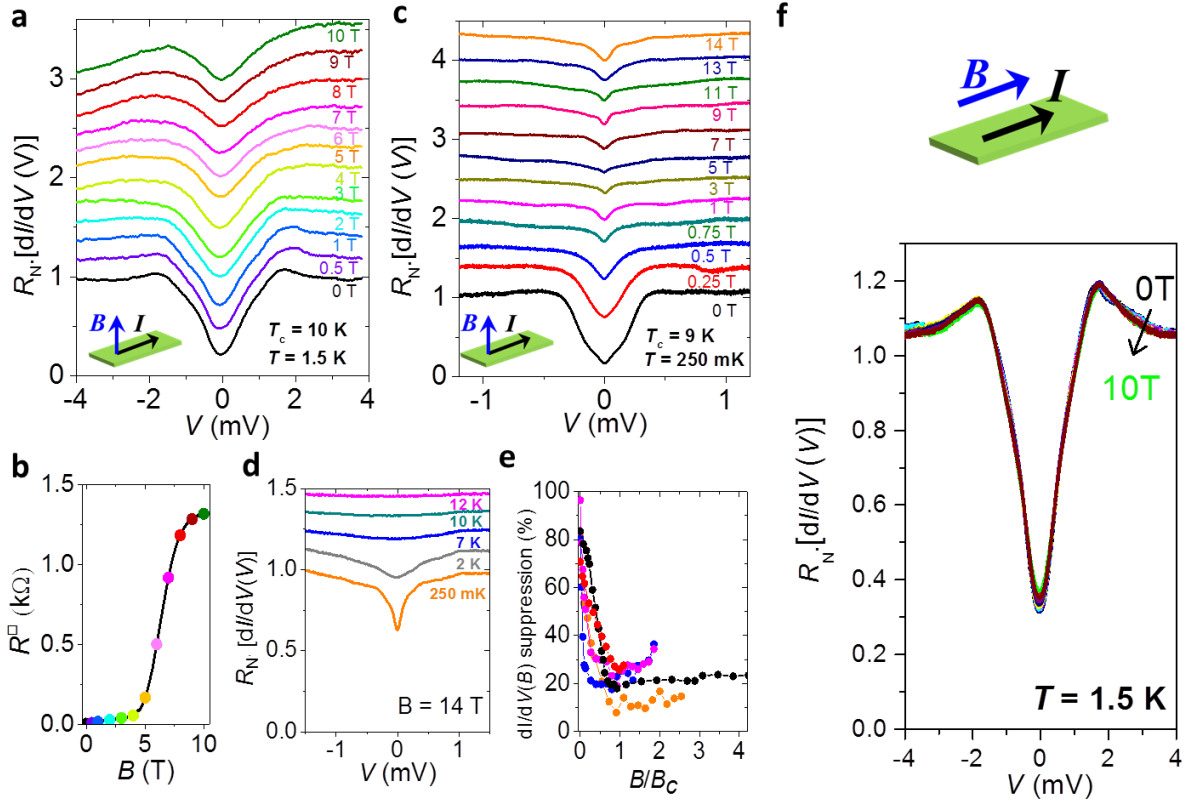

**Figure 4. Magnetic field dependence of the tunneling density of states.** **(a)** Evolution of the normalized differential conductance upon increasing the perpendicular magnetic field $B$, measured at $T = 1.5$ K on a device having $n = 1.5 \cdot 10^{14}$ cm$^{-2}$ and $T_c = 10$ K (curves offset for clarity). **(b)** Square resistance of the same device measured at $T = 1.5$ K as a function of $B$, showing a transition to the normal state resistance at $B = 9$ T. In **(a)** a pronounced suppression of the low-bias tunneling conductance is still clearly visible at $B = 10$ T, above the critical field $B_c$ inferred from transport measurements. **(c)** Evolution of the normalized differential conductance with increasing $B$ measured at $T = 250$ mK on another device ($n = 1.8 \cdot 10^{14}$ cm$^{-2}$ and $T_c = 9$ K; curves offset for clarity), for which the resistive transition gives $B_c = 9$ T. A pronounced DOS suppression persists at $B = 14$ T and only disappears for $T > T_c$ as shown in **(d)** by the $dI/dV$-$V$ measurements performed at $B = 14$ T. **(e)** Suppression in the zero-bias tunneling conductance relative to the normal state value plotted for five different devices as a function of $B/B_c$ (in all cases, for $T/T_c < 0.15$), showing that the suppression persists even for $B >> B_c$. **(f)** Normalized tunneling conductance as a function of bias measured at $T = 1.5$ K upon increasing the in-plane magnetic field from 0 to 10 T (for this device $n = 1.5 \cdot 10^{14}$ cm$^{-2}$ and $T_c = 10$ K). The in-plane magnetic field has no effect and leaves the DOS unchanged (the small variations around $V = 0$ can be accounted for by a few-degree misalignment of the field, resulting in a non-vanishing perpendicular component).

**Supplementary Information for the Article**

**Tunneling spectroscopy of gate-induced superconductivity in MoS$_2$**

Davide Costanzo[1] *, Haijing Zhang[1] *, Bojja Aditya Reddy[1], Helmuth Berger[2] and Alberto F. Morpurgo[1]**.

*These two authors contributed equally to this work*

**e-mail: alberto.morpurgo@unige.ch

This file includes:

Supplementary Text

Supplementary Fig. 1

Supplementary Fig. 2

Supplementary Fig. 3

Supplementary Fig. 4

Supplementary Fig. 5

Supplementary Fig. 6

Supplementary Fig. 7

In the main text we have presented all the key experimental observations made in our tunneling spectroscopy experiments and in this supplementary information we address a variety of issues that could not be discussed in the main text in full detail. As for what regards more technical aspects, we describe the device fabrication, and comment extensively on specific parts that proved to be important to improve the spatial homogeneity of carrier density. We also provide the technical specifications of the cryogenic systems employed in the experiments and discuss details of the electrical measurements. In the second part we address a number of points relevant for the interpretation of our data. We compare the tunneling conductance to calculations based on the DOS expected for a fully gapped superconductor and show that they cannot reproduce our observations. We also discuss –and exclude– some possible artifacts that could a priori affect the results of our tunneling spectroscopy, namely the effect of "pinholes" that could be present in the part of $MoS_2$ acting as tunnel barrier. Finally, we show I-V curves of the accumulation layer that indicate the 2D nature of the superconducting state.

**1) Device fabrication**

The devices discussed in the main text, used to perform tunneling spectroscopy on the gate-induced superconducting accumulation layer at the surface of $MoS_2$, were realized by a combination of conventional nano-fabrication techniques and of techniques that have been developed over the last few years to assemble so-called van der Waals heterostructures. The multilayer graphene (MLG) tunneling electrodes are defined by oxygen plasma etching of large flakes that are mechanically exfoliated from graphite onto a $Si/SiO_2$ substrate. Fig. S1a shows an optical microscope image of a MLG flake with a PMMA mask defining the electrode geometry; Fig. S1b shows the result after etching and removing the PMMA mask. The electrodes prepared in this way were then thermally annealed at 300 °C in an inert atmosphere to remove polymer residues from the MLG surface. Subsequently, $MoS_2$ multilayers were exfoliated from a bulk crystal and their thickness estimated by their optical contrast and using atomic force microscopy. Typically, large-area $MoS_2$ flakes consisting of 6-10 layers were selected and transferred onto the etched MLG strip with the aid of a micro-positioning system mounted under an optical microscope. This was done with either a PMMA membrane using a so-called wet-transfer technique[1], or by employing a PPC polymer film to pick up the flakes from a $Si/SiO_2$ substrate[2]. In either case, the resulting MLG-$MoS_2$ heterostructure was subsequently cleaned from polymer residues by annealing it at 300 °C in $H_2$/Ar (10:100 sccm). An optical microscope image of a device at this stage of the fabrication process is shown in Fig. S1c. Finally, two independent electron-beam lithography/metal evaporation/lift-off processes were done to attach metal contacts to both $MoS_2$ and MLG (Au electrodes were used for $MoS_2$ to minimize contact resistance; Ti/Au contacts were used for MLG, to maximize adhesion). In these same steps, a large Au pad acting as a gate electrode for the ionic liquid (DEME-TFSI, purchased from Kanto Corporation) was also deposited. The device was annealed once more at 200 °C for 2h in a flow of $H_2$/Ar (10:100 sccm) mixture, to reduce the contact resistance (at the high carrier density used in our investigations contact resistances were low and never posed particular problems).

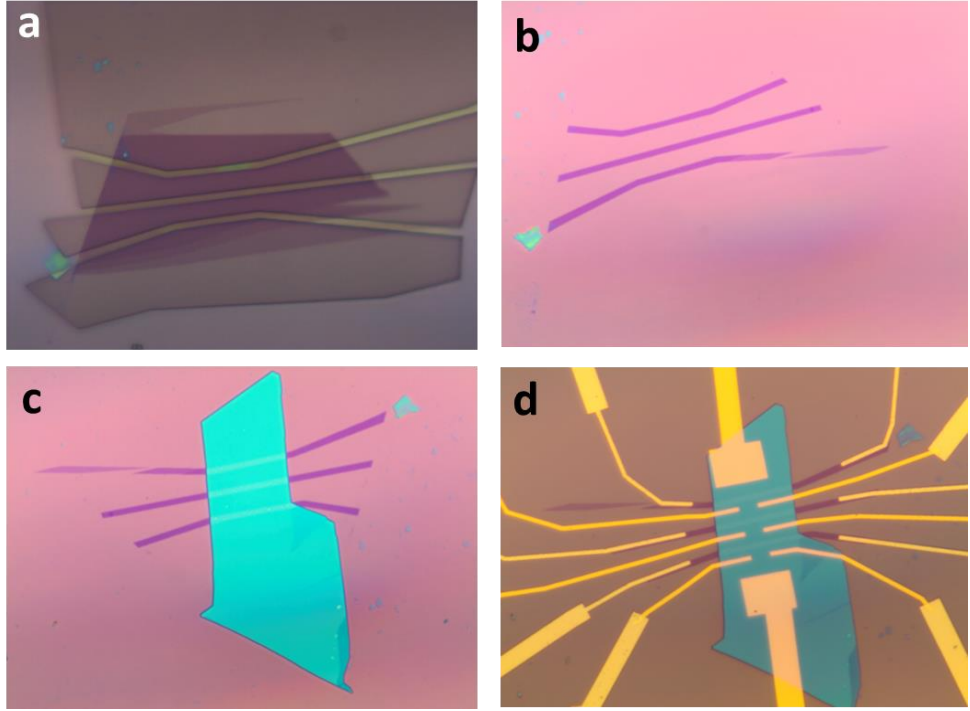

**Figure S1** | *Device fabrication*: **a**, optical microscope image of an exfoliated MLG flake on a Si/SiO$_2$ substrate covered by the PMMA mask defining the strips used as tunneling contacts. **b,** image of the same flake after etching in an oxygen plasma and removing the PMMA. Panel **c** and **d** show the device after having transferred the MoS$_2$ multilayer, and attached metal contacts to MoS$_2$ and MLG.

## 2) Improving the homogeneity of carrier density at low-temperature

In our earlier work on gate-induced superconductivity in WS$_2$ [3] and MoS$_2$ [4], we regularly found that at low temperature the gate-induced carrier density in different parts of a same device exhibited large inhomogeneity, which we could detect by using multiple pairs of probes to measure the Hall resistance in different locations. The phenomenon was caused by the local detaching of the frozen ionic liquid from the gated material, caused by mechanical stress generated upon cooling by the difference in the thermal expansion coefficient of the frozen liquid and of the substrate. Detaching of the frozen liquid also caused very large jumps in the resistance that we invariably observed when measuring the device resistance upon cooling from room temperature. Based on a multitude of transport measurements that we have done in the past on WS$_2$ and MoS$_2$ superconducting accumulation layers (Refs [3] and [4]), it appears that the liquid detaches along narrow lines, which can strongly affect the measured resistance (especially if they cross the entire width of the device) despite occupying a small fraction of the device surface. That is why it is seen that the temperature dependence of the resistance often depends strongly on the specific device measured (which is what other groups observe as well, as we gather from their publications; compare, e.g., Ref. [5], [6] and [7] ).

As our work on tunneling spectroscopy progressed, we managed to improve the situation. Specifically, we succeeded in decreasing the carrier density inhomogeneity and in minimizing the detaching of the frozen liquid.

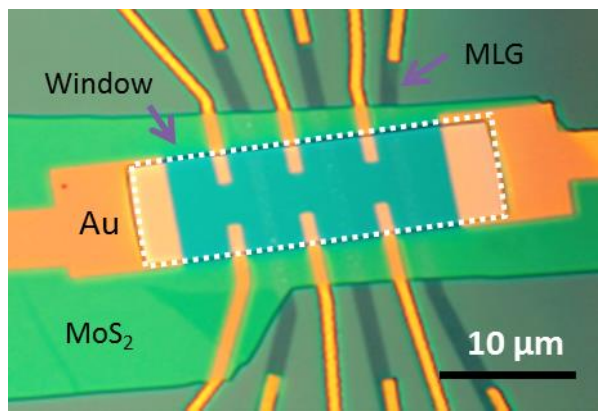

**Figure S2** | Optical microscope image of a device covered by a layer of PMMA, selectively removed only in the active region to be gated (i.e., a "PMMA window"), outlined by the white dashed line.

We found that two specific aspects of the device configuration and fabrication process have an important influence. The first concerns a so-called "PMMA window" that we used in the past to define the active part of the device (see Fig. S2). In the course of the present work we found that not using this PMMA window systematically improves the carrier density homogeneity and drastically reduces the resistance jumps that are seen upon cooling down the device.

There are two possible reasons why removing the PMMA window has such a beneficial effect. The first is that confining the frozen liquid and making the substrate surface non-planar is a dominant cause of mechanical stress, whose relaxation causes the frozen liquid to detach from the surface. The second is that using a PMMA window impedes any successive thermal annealing process (or solvent cleaning process). It follows that a much larger density of polymeric residues is present on devices in which a PMMA window is used. As these residues can strongly affect adhesion to the substrate, their presence may be largely responsible for the detachment of the frozen liquid. Indeed, we have additional evidence (see here below) that the presence of polymeric residues on the substrate surface does play an important role.

The second procedure that resulted in a drastic improvement in carrier density homogeneity is a so-called atomic force microscope (AFM) cleaning process[8, 9] of the $MoS_2$ surface. The process consists in gently sweeping the surface to be cleaned with an AFM tip in contact mode, for an extended period of time. As it has been repeatedly documented for different van der Waals materials, it allows dragging out of the AFM scanning field all (or at least the great majority of) polymer residues deposited during the device fabrication, thus effectively cleaning the surface in the scanned area. The process is illustrated by the AFM images shown in Fig. S3. We found that in devices for which this cleaning procedure was used the density homogeneity was further enhanced and that at most small resistance jumps were seen during the device cool down.

We can quantify the improvement in carrier density homogeneity resulting from these changes in the device configuration/fabrication process. In our earlier work on gate-induced superconductivity on $WS_2$ and $MoS_2$ variations in carrier density as large as a factor of 2 were rather regularly measured on gated flakes of ~20 μm linear dimension. In the present work, instead, we routinely succeeded in achieving fluctuations smaller than 20% on the same devices size: larger density fluctuations were observed only rarely. This result corresponds to a one-order-of-magnitude improvement.

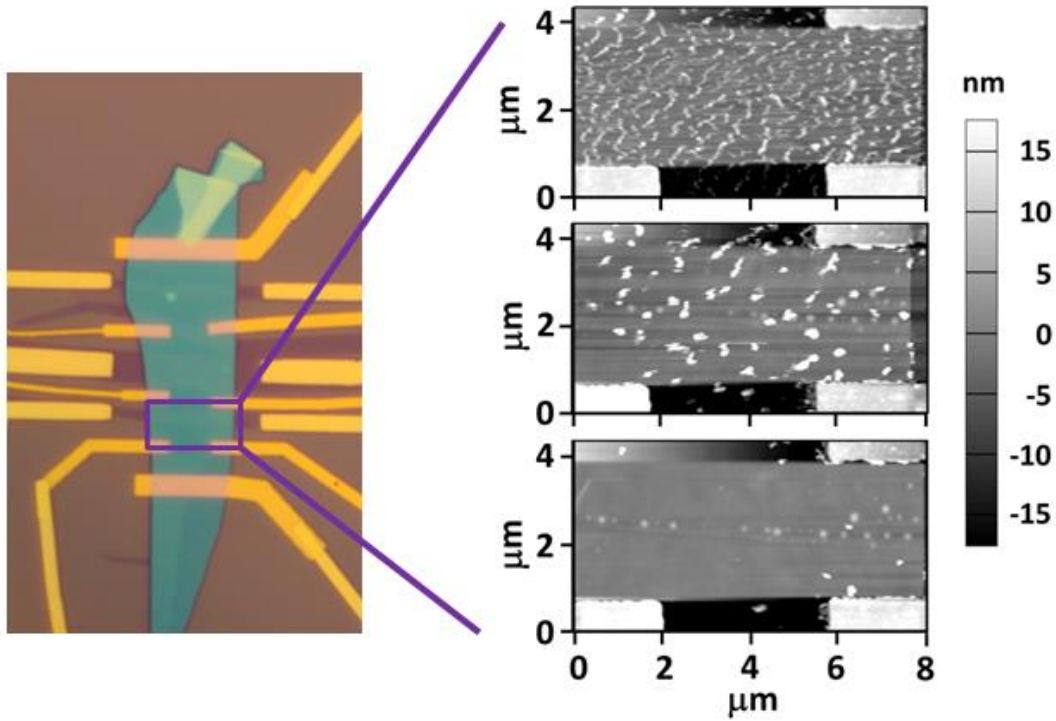

**Figure S3** | AFM cleaning is an effective and safe way to remove adsorbates and PMMA residues from MoS$_2$ flakes. It consists in a sequence of AFM scans in contact mode done with moderate force, which result in the accumulation of adsorbed residues on the side of the AFM scanning field. The left panel is an optical microscope image of one of our devices. AFM images of the rectangular area delimited by the purple line are shown in the right panel, at different stages of the cleaning process. Gradually, all adsorbates are clustered together to form larger particles that eventually are pushed out of the scanning field. In the bottom image in the right panel virtually all these larger particles have been removed and what is still visible are so-called "bubbles" that form in the region where MoS$_2$ is on top of one of the MLG strips (in the center of the image).

Having explained in detail the differences between devices with and without PMMA window for what regards carrier density homogeneity, it is worth discussing other information that can be extracted from the comparison of the general behavior of these devices. The key observation is that the result of the tunneling spectroscopy measurements were virtually identical in the two types of devices. Specifically, the collapse of the low-energy part of tunneling conductance at low energy is the same for both type of devices (that is, Fig. 3a and 3b include data measured on devices with and without PMMA window). A first implication is that this comparison indicates that inhomogeneity in carrier density is not responsible for the sub-gap DOS observed in the tunneling spectroscopy measurements. That is simply because the tunneling conductance is the same irrespective of the device type, but the density inhomogeneity is much larger in devices with PMMA window. A second direct and important implication is that the tunneling conductance that we have measured cannot be due to transport at the edges of the MoS$_2$ crystal, because in devices with PMMA window an extended region around the edges covered is not in contact with ionic liquid.

## 3) Cryogenic set-ups and electrical measurements

Different cryogenic systems have been used to perform the low-temperature electrical measurements discussed in the main text. The first is a variable temperature cryo-free Teslatron system (Oxford Instruments), equipped with a superconducting magnet allowing the application of fields up to 10 T, and reaching $T = 1.5$ K as lowest temperature. This system proved to be particularly convenient as it enables very slow and controlled device cooling. With a minimum temperature of $T = 1.5$ K this system gives enough energy resolution to perform spectroscopic measurements on devices at carrier densities close to optimal doping, having a critical temperature $T_c \sim 9\text{-}10$ K (i.e., not much lower than the maximum possible value for $MoS_2$). However, to have the energy resolution needed to investigate devices with a smaller $T_c$ –and hence a smaller superconducting gap $\Delta$– it is essential to perform measurements at sub-Kelvin temperatures. To this end, we used a Heliox $^3$He system (Oxford Instruments) operated in a He dewar equipped with a 14 T superconducting magnet, enabling us to reach temperatures as low as $T = 250$ mK. In this case achieving very slow cooling proved to be more complex. Nevertheless, cooling rates that were sufficiently slow not to cause problems with the frozen liquid could be achieved by pumping all the exchange gas out of the Heliox insert, and by gradually lowering the insert itself without directly immersing it into liquid He.

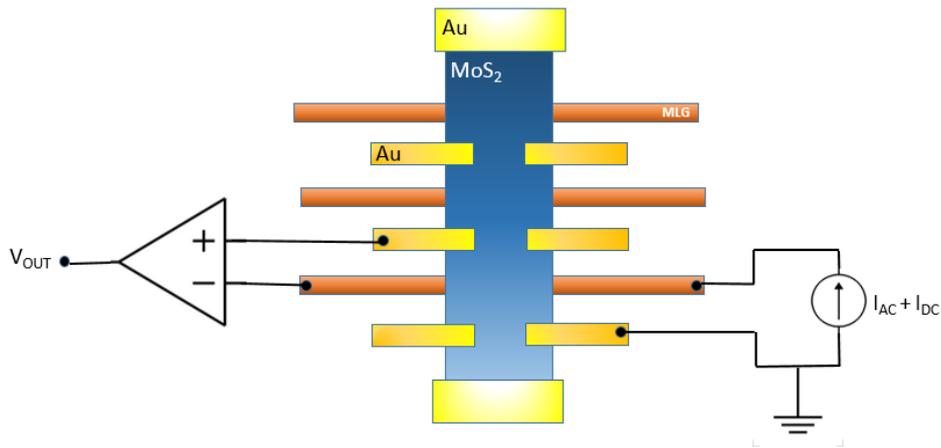

**Figure S4** | Schematic top-view of a device used for tunneling spectroscopy measurements with the four-terminal configuration used to measure the bias dependence tunneling conductance. This configuration removes from the measurements the possible influence of the series resistance due to either the MLG strip or to the interface between the gold electrodes and the $MoS_2$ accumulation layer.

In all cases, before cooling and/or performing any electrical measurement, devices were mounted in the cryostat and left under high vacuum for nearly a day, to ensure that all oxygen and humidity possibly present in the ionic liquid were removed. Only then a voltage was applied to the gate electrode and gradually increased through successive cycles, to reach the desired value (a procedure that also takes nearly a day). These precautions ensure the reversibility of the measurements and prevent chemical reactions between $MoS_2$ and either oxygen/water or adsorbates present at the surface. After the device was gate biased and cooled down, the accumulated electron density was inferred from Hall resistance measurements performed on the different pairs of Hall probes that we attached on each device (Hall measurements were done at a temperature higher than the superconducting critical temperature).

The differential conductance (*dI/dV-V*) measurements were performed in a current-biased four-terminal configuration, using low-noise home-made electronics to add an ac and a dc current and to amplify the resulting voltage signal (see fig. S4). The ac part of the amplified voltage was measured synchronously with a Stanford SR830 lock-in amplifier; an Agilent 34401a digital multimeter was used for the DC component. The voltage to the gate electrode was applied using a K2400 source-meter unit. Care was taken to apply the current and measure the voltage in a four-terminal configuration such as the one illustrated in Fig. S4, to eliminate from the measurements the effect of any series resistance possibly coming from either the MLG strip or to the interface between the gold electrodes and the $MoS_2$ accumulation layer.

For the different devices investigated, the low-temperature tunneling resistance (i.e., the resistance between the MLG strips and the gate-induced accumulation layer at the surface of $MoS_2$) ranged between several kOhms and approximately 100 kOhms, depending mainly on the thickness of the $MoS_2$ layer used and the area of the tunneling contact, with most typical values of the order of 10-20 kOhms. The bias-dependence differential resistance did not depend on the absolute value of the tunneling resistance.

### 4) Comparison of the tunneling conductance with predictions for a fully gapped superconductor.

In the main text, we claimed that a fully gapped superconducting state cannot reproduce the data. Here we provide support to this statement by comparing the measured bias-dependent tunneling conductance with different models for conventional s-wave fully gapped superconductors, in the ideal homogenous case (part **4a**), taking into account the presence of carrier density inhomogeneities (part **4b**), or including the effect of finite quasi-particle lifetime (part **4c**).

In all cases, the differential tunneling conductance is calculated directly from the conventional expression of the tunneling current[10] :

$$I(V) = \frac{1}{eR_N} \int_{-\infty}^{\infty} \frac{N_S(\varepsilon)}{N(0)} \times [f(\varepsilon, T) - f(\varepsilon + eV, T)] \cdot d\varepsilon$$

where $N_S(\varepsilon)$ is the DOS in the superconducting state under consideration, $N(0)$ the DOS at the Fermi energy in the normal state, $f(E)$ and $f(E + eV)$ are the Fermi-Dirac distributions in the electrodes on opposite sides of the tunnel barrier. Expanding $f(E + eV)$ for small values of V (since $eV \ll E_F$) and differentiating, we obtain:

$$\frac{dI}{dV}(V) = \frac{1}{eR_N} \int_{-\infty}^{\infty} \frac{N_S(\varepsilon)}{N(0)} \times \frac{-\partial f}{\partial V}(\varepsilon + eV, T) \cdot d\varepsilon \quad (S1)$$

**a) Fully gapped state**

We start by considering a conventional s-wave superconductor, whose DOS is $N_S(\varepsilon) = 0$ if $|\varepsilon| < \Delta$ and $N_S(\varepsilon) = N(0)\frac{|\varepsilon|}{\sqrt{\varepsilon^2-\Delta^2}}$ if $|\varepsilon| > \Delta$. With regard to the low-energy sub-gap tunneling conductance, identical considerations would apply to a two band superconductor with the same sign of the gap (such as $MgB_2$).

To compare with experiments, we consider the case of the device whose data are shown in Fig. 2b (and reproduced in Fig. S6a), for which the gap $\Delta \cong 1.5$ meV. At a temperature of 1.5 K ($kT = 0.13$ meV), the tunneling conductance calculated using Equation S1 is plotted in Fig. S5a. It is clear that the pronounced suppression of sub-gap tunneling conductance throughout a broad bias interval is not in agreement with the data. Therefore, considering a plain fully gapped state (such as in a s-wave superconductor) cannot explain our observations.

**b) Fully gapped state with charge density inhomogeneity**

Inhomogeneity in carrier density could lead to an enhanced tunneling conductance if it was so large to cause the presence of non-superconducting regions (or of regions having a superconducting gap comparable to the lowest $T$ reached in the experiments) in the accumulation layer at the $MoS_2$ surface. Here we show that an inhomogeneity in carrier density cannot reproduce our experimental results even if we assume that this inhomogeneity is much larger than that extracted from measurements of the Hall resistance in different parts of the device[1].

Our data (Fig. 3c of the main text) show that the superconducting gap decreases from about $\Delta = 2$ meV at $n \sim 10^{14}$ cm$^{-2}$ to about $\Delta = 0.35$ meV for $n \sim 5 \cdot 10^{14}$ cm$^{-2}$, and a linear interpolation between these values provide a good overall estimate for $\Delta(n)$ (i.e. the value of $\Delta$ for any value of $n$ in this range). We estimate the effect of inhomogeneity in carrier density if we consider that in large-area tunnel junction such as the ones used in our experiments, regions with different $n$ (and a correspondingly different gap $\Delta$) contribute "in parallel" to the total measured tunneling conductance. In other words, the tunneling conductance of a fully gapped superconducting state in the presence of carrier density inhomogeneity can be calculated using Equation S1 and averaging it over the distributions of values of the gap $\Delta$ corresponding to the assumed density inhomogeneity.

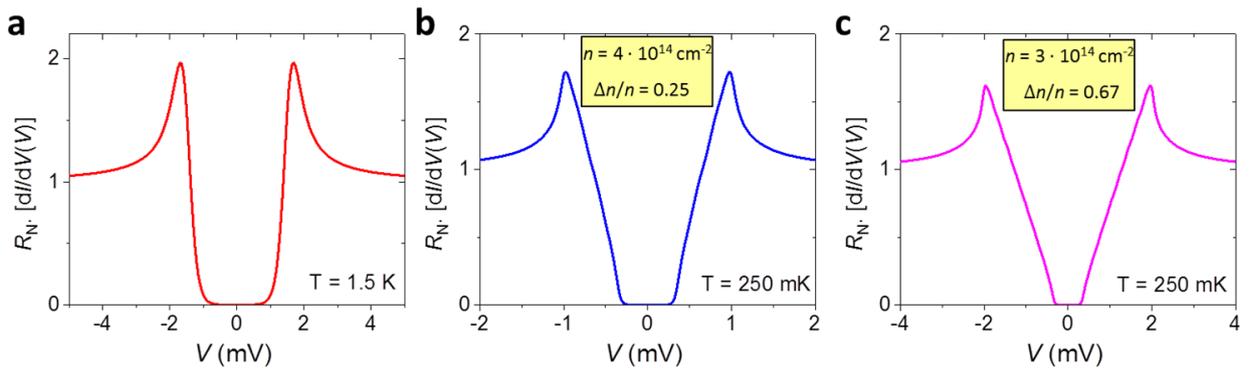

**Figure S5** | Calculated bias dependence of the tunneling conductance expected for a fully gapped superconductor. Panel **a** shows the case expected for a homogeneous superconductor with $\Delta = 1.5$ meV and T =1.5 K. The fully suppressed conductance is inconsistent with experimental obervations. Panel **b** and **c** include the effect of carrier density inhomogeneity that makes the gap position dependent. For the density dependence of the gap we used the values of $\Delta(n)$ inferred from the measurements. The two curves correspond to two specific cases: the left panel to an average carrier denity $n = 4 \cdot 10^{14}$/cm$^2$ with relative fluctuations of ±25%; in the right panel the average carrier density is $n = 3 \cdot 10^{14}$/cm$^2$ and the relative fluctuations are chosen to be ±67% (much larger than what is found in the measurements). In both cases the temperature is chosen to be 250 mK. It is seen that despite the large density fluctuations, a virtually vanishing tunneling conductance is found for a broad interval of applied voltage around $V=0$, at odds with the experimental observations.

---

[1] We present this analysis as an additional confirmation, because –as we discussed in the main text– the scaling of the low-bias tunneling conductance in all devices when data are plotted versus reduced variables already provides clear evidence that electron density inhomogeneity does not play a relevant role at low energy.

We have done the calculation for experimentally relevant conditions and seen that in all cases at the temperature of the measurements (250 mK) an exponential suppression should be observed for bias smaller than 0.3 meV. The results of the calculations are illustrated in Fig. S5 with two specific cases: $n = 3 \cdot 10^{14}$ cm$^{-2} \pm 2 \cdot 10^{14}$ cm$^{-2}$ (i.e., $\pm 67\%$ fluctuations in carrier density, a much more "pessimistic" estimate than the value inferred from Hall resistance measurements) and $n = 4 \cdot 10^{14}$ cm$^{-2} \pm 1 \cdot 10^{14}$ cm$^{-2}$ (corresponding to $\pm 25\%$ fluctuations). We see that the density fluctuations smear the characteristic side lobes present at the gap, and lead to a more gradual decrease of the tunneling conductance for bias a little smaller than the gap, but do not lead to the finite low-bias tunneling conductance observed in the experiments.

In short: charge density inhomogeneity affects strongly the tunneling conductance for large bias, $eV \sim \Delta$, in agreement with what we see experimentally. However, in a fully gapped s-wave state with realistic parameter, density inhomogeneity cannot cause the behavior that we observe in the experiments, namely a tunneling conductance that depends linearly on bias and vanishes for $V = 0$ at sufficiently low temperature, irrespective of the value of $n$.

### c) Fully gapped superconductor with finite quasiparticle lifetime

An additional possibility to consider is a fully gapped superconductor with a finite quasiparticle lifetime, a situation which is often considered in the analysis of tunneling experiments and that is described phenomenologically by a parametrization of the DOS due to Dynes [11], namely $N_S(\varepsilon) = N(0) \times Re\left(\frac{|\varepsilon+i\Gamma|}{\sqrt{(\varepsilon+i\Gamma)^2-\Delta^2}}\right)$. Here, $\Gamma$ is the energy corresponding to the quasiparticle decay rate: for this parametrization to be physically meaningful, $\Gamma$ should be much smaller than $\Delta$ and should not increase upon lowering temperature

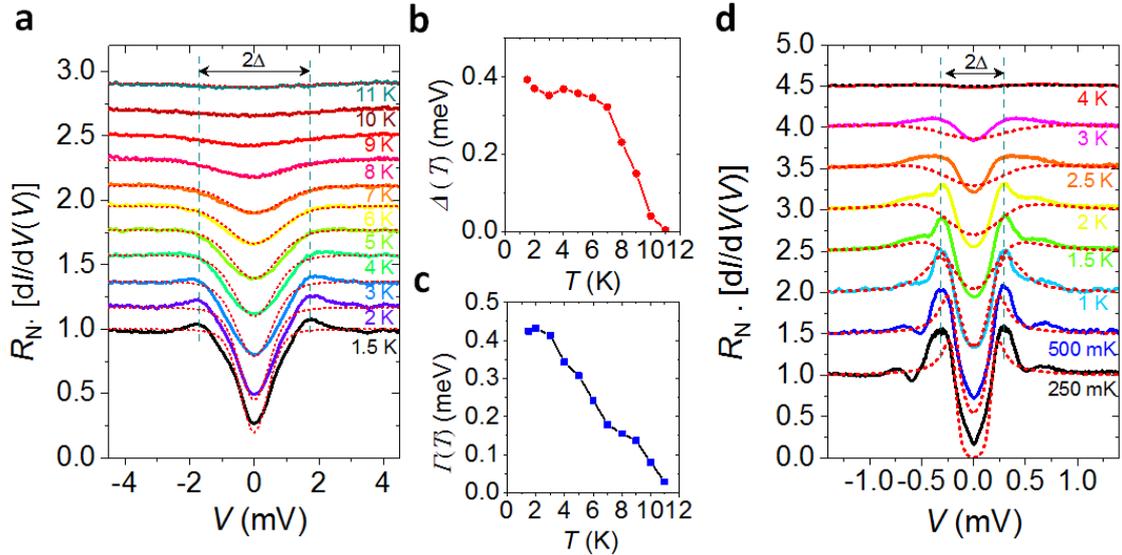

**Figure S6** The dotted lines in **a** and **d** are an attempt to reproduce experimental data with the Dynes parametrization of the DOS, to include finite quasiparticle lifetime effects. In **a**, where the agreement with data may seem acceptable, the curves have been calculated with the T-dependent values of $\Delta$ and $\Gamma$ shown in **b** and **c**, which are unphysical ($\Delta \sim \Gamma$ and $\Gamma$ increases upon lowering T). In **d** a constant value of $\Gamma$ has been used.

The dotted lines in Fig. S6 a and d show an attempt to use the Dynes function to reproduce the experimental data (the same shown in the main text in Fig. 2b and 2d). In Fig. S6a, the coherence peaks in the experimental data –even though shallow– are visible at low temperature; in Fig. S6d they are considerably more pronounced. One may think that in Fig. S6a the dotted line reproduces the data and their evolution with temperature in a reasonable way (albeit no sign of coherence peak is present in the calculated curves). However, by looking at Fig. S6 b and c –which shows the temperature dependence of the values of $\Delta$ and $\Gamma$ used to calculate the theoretical in Fig. S6a – it is apparent that the choice of parameter is entirely unphysical: $\Gamma$ increases with lowering temperature and $\Delta \sim \Gamma$. If we attempt to reproduce data of Fig. S6d in which the coherence peaks are more pronounced (such as in Fig. S6d) the situation is even worse, and we are forced to conclude that it is impossible to capture simultaneously the behavior observed experimentally for $eV \sim \Delta$ and for $eV \ll \Delta$. If the value of $\Gamma$ is chosen so that the coherence peaks are not entirely washed out, the low bias suppression of the tunneling conductance is much more pronounced than in the experiments; if $\Gamma$ is chosen to reproduce the low-bias part of the curve, no sign of coherence peaks remains. Therefore, a fully gapped superconducting state does not reproduce the tunneling conductance even if we allow for finite lifetime quasiparticle effects.

## 5) Excluding the effect of defects in the tunnel barrier

From the normal state value of the measured tunneling conductance, using the known electronic properties of graphite, we can easily estimate the average transmission $\langle T \rangle$ of electronic modes responsible for tunneling from the MLG electrode to the $MoS_2$ accumulation layer. We find $\langle T \rangle \sim 10^{-3} - 3 \cdot 10^{-5}$ depending on the specific device considered. Since $\langle T \rangle \ll 1$ in all cases, our devices are always in the tunneling regime, as we assumed throughout the discussion in the main text. It could however happen that, owing to the presence of a defect in the $MoS_2$ tunnel barrier (a pinhole) one or a few modes propagating from the MLG electrode to the $MoS_2$ accumulation layer have a transmission probability $T$ much larger than the average, i.e. $T \cong 1$. Such modes would then give an entirely non-negligible contribution to the measured conductance that –having $T \cong 1$– would not be suppressed at sub-gap voltages (in fact it would likely be enhanced, because of Andreev reflection). One should then consider the possibility that the low-bias conductance that we measure originates from such an effect, and not from the absence of a full gap.

Invoking pinholes in the tunnel barrier is not consistent with the observation that the low-bias tunneling conductance is the same in all devices when plotted as a function of $eV/\Delta$: defects such as pinholes are device specific and would not lead to precisely the same behavior in different devices. The effect of pinholes can also be excluded because the bias dependence of the conductance that we measure is the same irrespective of the normal state resistance of the junction considered. Since the normal state resistance of our devices is not much smaller than the quantum of resistance $R_Q = \frac{h}{e^2}$, the presence of even an individual highly transmitting mode would imply that this single mode dominates transport also in the normal state (i.e., not only in the sub-gap regime). Such a conclusion would not be consistent with the observed pronounced suppression of the low-bias differential conductance, which vanishes at low energy. We can be even more specific, because some of the devices that we measured have a normal state resistance $R_N \cong 100$ k$\Omega$ (or even somewhat larger), i.e. substantially higher than $R_Q$. In these devices, even if all the measured conductance was attributed to an individual highly-transmitted mode caused by the presence of a pinhole, we would be forced to conclude that the transmission probability of that mode is at most $T \sim 0.2$ (or less). A transmission probability $T \sim 0.2$ is already sufficiently low to be in the tunneling regime. That is: for the highest resistance devices no matter what assumption is made, we have to conclude that transport is in the tunneling regime.

In summary, the behavior of the low-bias conductance that we observe experimentally cannot be accounted for by defects in the tunnel barrier causing the presence of highly transmitted modes propagating from MLG to the accumulation layer at the surface of $MoS_2$.

## 6) Manifestation of Berezinsk*ĭi*-Kosterlitz-Thouless superconductivity near $T_c$

Due to the small electrostatic screening length at $n \sim 10^{14}$ cm$^{-2}$ (estimated to be approximately 1 nm) the superconducting state induced at the surface of ionic liquid gated $MoS_2$ transistors is expected to have a two-dimensional (2D) nature irrespective of the thickness of the $MoS_2$ crystal used. In 2D, although long-range order of the superconducting order parameter is not possible due to thermal fluctuations, quasi-long-range order can still occur, due to the so-called Berezinsk*ĭi*-Kosterlitz-Thouless (BKT) mechanism. A BKT transition is characterized by a critical temperature ($T_{BKT}$, lower than the mean field $T_c$) below which phase stiffness become finite and a supercurrent can be measured, despite the absence of long-range order.

A BKT regime has different peculiar experimental manifestations. One of them is the power law dependence of the I-V characteristics, i.e. $V \propto I^\alpha$, where $\alpha = 1$ for $T$ well above $T_{BKT}$, $\alpha = 3$ for $T = T_{BKT}$, and $\alpha > 3$ for $T < T_{BKT}$. This behavior has indeed been observed in our devices for temperatures below the onset temperature of superconductivity. This is shown in Fig. S7a, where I-V curves measured on a $MoS_2$ FET are plotted in a double logarithmic scale: the clearly visible linear regime extending over several orders of magnitude is indicative of a power law relation between $I$ and $V$. The temperature dependence of the exponent, which corresponds to the slope $\alpha$ in the log-log plot, is shown in the inset of Fig. S7a. It can be seen that $\alpha$ increases upon lowering the temperature, as expected; $\alpha = 3$ occurs for $T$=7.4 K, providing a first estimate of $T_{BKT}$.

The BKT nature of the transition is confirmed by looking at the temperature dependence of the resistance close to $T_{BKT}$. BKT theory predicts $R = R_0 \exp(-bt^{-1/2})$ ($R_0$ and $b$ are material dependent parameters, and t = $T/T_{BKT} - 1$), from which it is expected that $[d(\ln R)/dT]^{-2/3}$ varies linearly with $T$ above $T_{BKT}$, and that it extrapolates at 0 for $T = T_{BKT}$. Fig. S7b shows that this is indeed the case. From the data $T_{BKT}$ is estimated to be 7.4 K, consistently with the value extracted from the exponent of the power law relation of the I-V curves. The consistency of these two measurements, confirms that the experimental behavior observed for $T$ close to $T_c$ conforms to that of 2D BKT superconductivity.

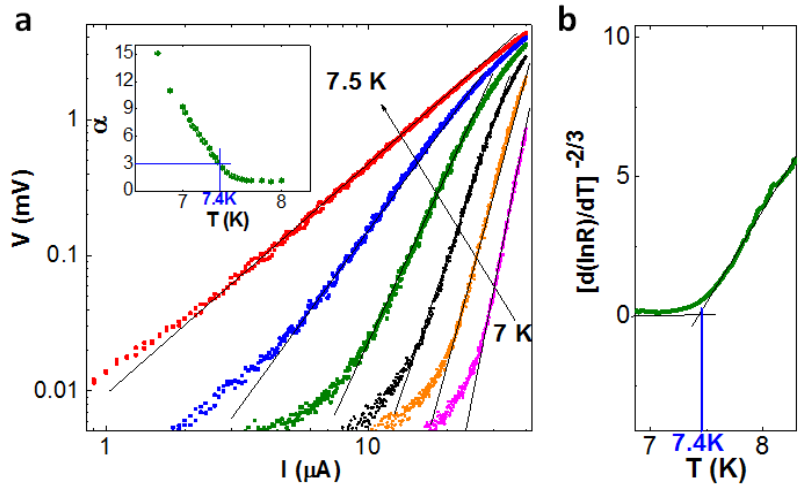

**Figure S7 | Berezinsk*ĭ*i-Kosterlitz-Thouless transition in superconducting MoS$_2$. a,** Double-logarithmic plot of the *I-V* characteristics of a gate-induced superconducting MoS$_2$ device at different temperatures near the transition. The observed linear behavior implies that $V \propto I^\alpha$. The solid black lines are linear fits to extract the exponent $\alpha$. The inset shows that temperature dependence of $\alpha$. $T_{BKT}$ corresponds to the value of $T$ for which $\alpha = 3$, from which we infer $T_{BKT}$ = 7.4 K) **b,** Linear behavior of $[d(\ln R)/dT]^{-2/3}$ as a function of temperature. The extrapolation to zero (black solid line) provides an independent estimate of $T_{BKT}$, which also gives 7.4 K.